\documentclass[fleqn,usenatbib]{mnras}

\usepackage{newtxtext,newtxmath}

\usepackage[T1]{fontenc}

\DeclareRobustCommand{\VAN}[3]{#2}
\let\VANthebibliography\thebibliography
\def\thebibliography{\DeclareRobustCommand{\VAN}[3]{##3}\VANthebibliography}

\usepackage{graphicx}	
\usepackage{amsmath}	

\title[Shapes of ICL and DM haloes]{Intracluster light is a close tracer of the dark matter halo shape}

\author[A. Fernandez et. al.]{
Adela Fernandez,$^{1}$\thanks{E-mail: ppyaf@nottingham.ac.uk}
Yannick M.~Bah\'{e},$^{1,2}$
Nina A.~Hatch,$^{1}$
Joseph Butler,$^{1}$
Tutku Kolcu,$^{1}$\newauthor
~Garreth Martin,$^{1}$
and Mireia Montes$^{3}$
\\
$^{1}$School of Physics and Astronomy, University of Nottingham, University Park, Nottingham NG7 2RD, UK\\
$^{2}$Laboratoire d'Astrophysique, \'{E}cole Polytechnique F\'{e}d\'{e}rale de Lausanne (EPFL), Observatoire de Sauverny, 1290 Versoix, Switzerland\\
$^{3}$Institute of Space Sciences (ICE, CSIC), Campus UAB, Carrer de Can Magrans, s/n, 08193 Barcelona, Spain
}

\date{Accepted XXX. Received YYY; in original form ZZZ}
\pubyear{\the\year{}}

\begin{document}
\label{firstpage}
\pagerange{\pageref{firstpage}--\pageref{lastpage}}
\maketitle

\begin{abstract}
We investigate whether the intracluster light (ICL) can serve as a reliable tracer of the shape of the underlying dark matter (DM) haloes in galaxy clusters. Using the cosmological Hydrangea cluster simulations, we measure the 3D and projected shapes of both components with a shape tensor computed in concentric ellipsoidal shells, out to the virial radius $R_\mathrm{200c}$ for each cluster. The ICL and DM are closely aligned, with their major axes typically offset from each other by $\lesssim$10\,degrees. Their axis ratios also match closely, with a typical difference of only $\approx\! 0.07$ for both the major-to-minor and major-to-intermediate axes, the DM being slightly rounder than the ICL. These trends are consistent across 2D and 3D measurements and agree well with results from isophotal fitting of mock images. In detail, the axis ratio offset is sensitive to the method used to remove satellites, and may also depend on the choice of subgrid physics models. We demonstrate that the ICL traces the DM shape better than the distribution of satellite galaxies, which exhibits larger scatter in the axis ratio and misalignment angle and is overall more elliptical. Together, these results indicate that the ICL can act as a useful proxy for DM halo ellipticity and orientation.
\end{abstract}




\section{Introduction}

There is strong evidence that the majority of the mass in the Universe is not baryonic, but is instead provided by a poorly understood component commonly referred to as ``dark matter'' (DM). While its nature remains unknown, a range of phenomenological models have been proposed to describe its large-scale behaviour. The most widely accepted theory is the cold dark matter model \citep[CDM;][]{davis_evolution_1985}, where  DM behaves as non-relativistic and collisionless particles. This model provides a robust description of structure formation in the Universe and reproduces the cosmic microwave background anisotropy power spectrum with high precision (e.g.~\citealt{Planck2018VI}). While CDM provides an excellent match to large-scale structure, several potential small-scale tensions remain, such as the missing satellite \citep{Moore1999ApJ...524L..19M, Klypin1999ApJ...522...82K}, plane of satellites (e.g.~\citealt{Ibata2013}), and core-cusp problems \citep{Moore1994Natur.370..629M, deBlok2010AdAst2010E...5D}, and variants thereof (e.g.~\citealt{BoylanKolchin2011, Mueller2024}). These motivate the continued exploration of alternative DM models. Examples of such alternatives include warm dark matter (WDM; e.g.~\citealt{colin2000}) where particles possess non-negligible thermal velocities that lead to the suppression of small-scale clustering, and self-interacting dark matter (SIDM; \citealt{Spergel2000}), where DM particles interact with each other also through some non-gravitational force and hence behave as a collisional fluid.

Since DM dominates the gravitational collapse of structures, the assembly history of DM haloes, and therefore their shapes, depends on the nature of DM itself. For example, CDM haloes are typically triaxial and elongated \citep{dubinski_1991, jing_triaxial_2002}, whereas in SIDM the self-interaction leads to scattering of DM particles and hence rounder halo shapes \citep{peter_cosmological_2013}. In WDM cosmologies, substructure is suppressed, leading to fewer satellite haloes \citep{lovell_haloes_2012}, and slightly rounder inner density distributions \citep{maccio_inner_2013}. Consequently, the shape of DM haloes in observations may enable us to probe the nature of DM itself.

DM haloes span a very wide mass range, with galaxy clusters at the extreme high-mass end. In these systems, the influence of baryonic processes is less significant compared to smaller systems such as galaxies or groups (e.g.~\citealt{McCarthy2010MNRAS.406..822M, Ayromlou2023MNRAS.524.5391A}); the total mass and DM radial density profiles therefore agree closely between simulations with and without baryons, outside the central galaxy \citep{schaller_baryon_2015}. Assuming that the same holds for their shapes, clusters can be regarded as powerful laboratories for probing the properties of DM through the shapes of their haloes.

Measuring any property of DM haloes is, however, challenging because DM cannot be observed directly and must instead be inferred through observational proxies. X-ray emission from the hot intracluster medium (ICM) has traditionally been used as for this purpose \citep[e.g.][]{eyles_distribution_1991}. But unlike DM, this gas is collisional and is also influenced by baryonic processes, such as active galactic nuclei (AGN) feedback, turbulence, and shocks. These influences make it an unreliable tracer of the DM in clusters, particularly within unrelaxed clusters \citep{Smith2016MNRAS.456L..74S}. Gravitational lensing provides an alternative probe of the cluster mass distribution, which to first order traces the underlying DM halo directly \citep{clowe2006}. However, the lensing signal is weakest in the cluster outskirts, precisely in the regions of greatest interest for measuring the DM shape, where the influence of baryons is smallest. Consequently, applying this method to individual clusters -- a necessity to study shapes without a priori knowledge of a preferred orientation for stacking -- is limited to very massive systems with deep imaging \citep[e.g.][]{Diego2025arXiv250708545D}. The projected distribution of satellite galaxies offers another proxy \citep{shin2018}, though their sparse sampling can bias the inferred shapes toward higher ellipticities \citep{herle_unbiased_2025}. 

Given the limitations of these traditional proxies, intracluster light (ICL) has emerged as a promising luminous tracer of DM. ICL, first observed by \citet{zwicky_coma_1951}, is the light emitted by stars not bound to a satellite galaxy, which instead move freely in the gravitational potential of the cluster (see \citealt{Contini2021} and \citealt{montes2022NatAs...6..308M} for recent reviews). These stars are expected to behave as effectively collisionless particles due to their high velocities and the low stellar density in the intracluster space. Their dynamics are therefore governed primarily by the gravitational potential of the cluster \citep{rudick_formation_2006}. Most intracluster stars are thought to originate from objects that also contribute significantly to the assembly of the cluster's DM halo \citep[e.g.][]{Brown2024MNRAS.534..431B}. Consequently, the ICL can be expected to trace the shape of this DM halo, and potentially more closely than the ICM or satellites.

The surface brightness of the ICL decreases steeply with distance from the cluster centre \citep[e.g.][]{Zibetti2005MNRAS.358..949Z, Krick2006}, meaning that it has so far mostly been studied within $\lesssim 200$ kpc from the brightest cluster galaxy (BCG). However, the arrival of wide-field survey telescopes coupled with advances in data processing techniques, such as improved flat fielding and removal of backgrounds from the instrument or zodiacal light (e.g.~\citealt{Borlaff2019}; \citealt{Montes2021}; \citealt{ahad2025A&A...702A.271A}), now allow us to observe the ICL throughout the full halo. For example, \citet{zhang_dark_2024} stacked ground-based images of over 4000 clusters from the Dark Energy Survey (DES) to measure the ICL radial profile to 1\,Mpc. \citet{gonzalez_discovery_2021} measured the ICL surface brightness profile of an individual system, the Frontier Fields cluster MACS J1149.5+2223, out to 2\,Mpc. Their study builds on previous work with the Frontier Fields, which provided detailed ICL measurements for 6 very massive clusters using deep Hubble Space Telescope images \citep{montes_intracluster_2018}. More recently, {\it Euclid} Early Release Observations have detected the ICL in the Perseus and Abell 2390 clusters out to 600 kpc \citep{kluge_euclid_2025, ellien_euclid_2025}. Upcoming surveys such as the full \textit{Euclid} Wide Survey \citep{2011arXiv1110.3193L, scaramella2022} and the Vera C. Rubin Observatory's Legacy Survey of Space and Time (LSST; \citealt{Ivezic2019}) will substantially advance our observational capabilities, enabling similar analyses for thousands of clusters \citep{bellhouse2025A&A...698A..14E} and extending both the statistical power and depth of ICL studies.

First analyses have shown that the angular distribution of the ICL, i.e. its shape, indeed broadly traces that of the DM, even though the radial density profiles of the two components differ significantly from each other \citep{Pillepich2018, alonsoasensio_intracluster_2020, contreras-santos_characterising_2024, butler_intracluster_2025}. For example, \citet{montes_intracluster_2019} compared the mass distribution inferred from strong lensing to the ICL distribution in the Frontier Fields clusters. They found that the ICL isocontours trace the mass contours more closely than those of the hot X-ray emitting gas, up to radii of 140\,kpc, indicating that the ICL is a more reliable tracer of the underlying DM. Nevertheless, observational measurements of the DM shape are challenging: the 2D projection inherent to lensing, together with the contributions from foreground and background substructure, can bias inferred halo properties (e.g.~\citealt{bahe_mock_2012, Debackere2022, Vecchi2025}). Likewise, the ICL shape measurements may be sensitive to observational systematics such as masking of satellite galaxies, PSF effects, and uncertainties in sky or background subtraction.

In light of these limitations, it is instructive to examine what simulations predict about the agreement between the shapes of the two components. Simulations provide the only means to identify and measure the DM distribution accurately and across the full extent of the cluster halo. Moreover, they allow a direct separation between DM and baryons and provide their full 3D distributions, thus allowing a clean and physically motivated distinction between the smooth DM and ICL background haloes from satellite substructures. Despite their inherent nature as simplified models, the formation of both ICL and DM haloes is followed self-consistently as both are governed, to leading order, only by gravity whose treatment in modern simulations is numerically robust. Simulation predictions about the ICL are therefore not strongly affected by uncertain subgrid physics choices (see \citealt{brough_preparing_2024}.) and allow physically meaningful insight into the real Universe.

Previous simulation studies have compared the 2D spatial distributions of the ICL and DM shapes; for example those by \citet{alonsoasensio_intracluster_2020} and \citet{yoo2024ApJ...965..145Y}, who both found good agreement between the two components. Their approaches focus on the similarity of isodensity contours, using metrics such as the Modified Hausdorff Distance and the Weighted Overlap Coefficient. However, in the literature on DM halo shapes and in studies testing alternative DM models, halo morphology is typically characterised by modelling haloes as ellipsoids or ellipses and analysing their axis ratios (e.g.~\citealt{despali_like_2014, lau_correlations_2021}). Therefore, there is a need to quantify the difference in shape between the ICL and DM using the same ellipsoidal metrics, to characterise their intrinsic shapes and evaluate the extent to which ICL can be used as a tracer of the underlying halo shape.

In this paper, we investigate the reliability of the ICL as a luminous tracer of the shape of DM haloes in the Hydrangea simulations \citep{bahe_hydrangea_2017}, a suite of 24 cosmological hydrodynamical cluster zoom-in simulations based on the EAGLE model \citep{schaye_eagle_2015}. We quantify and compare the shapes of the two components in terms of their position angles and axis ratios, in both 2D and 3D. In addition, we compare our results to shapes measured using isophotal fitting of mock images, as is typically done in observations, and to those measured from the projected distribution of satellite galaxies. We do not compare our results to measurements of gas particle shapes, as these are strongly affected by the more uncertain aspects of sub-grid models. This comparison therefore lies beyond the scope of this paper, although it would be an interesting avenue for future work.

The remainder of this paper is structured as follows. In Section\,\ref{sec: sims and BCG+ICL definition}, we summarize the key aspects of the Hydrangea simulations and the ICL definition used in this work, and describe our shape measurement method. In Section\,\ref{sec: ICL DM shapes}, we present the results for DM and ICL in 2D and 3D and compare to equivalent measurements of the ICL shape from isocontour fitting of mock images. In Section\,\ref{sec:sim_comparison} we assess the numerical robustness of our results by comparing different satellite removal methods and a different simulation (Horizon-AGN; \citealt{dubois_dancing_2014}), and in Section\,\ref{sec:other_tracers} we compare our results to the satellite galaxy distribution. Our results are discussed in the context of other works in Section\,\ref{sec:discussion}, before summarizing our main conclusions in Section\,\ref{sec:conclusions}. Unless otherwise specified, all measurements are quoted in proper units and do not contain additional factors of `little $h$'.

\section{Simulations, ICL definition and Shape Measurements}
\label{sec: sims and BCG+ICL definition}
\subsection{The Hydrangea simulations}
Accurately modelling the ICL requires large simulation volumes to obtain statistically representative cluster samples, but also high spatial and mass resolution to model the formation of this diffuse stellar component in a numerically robust way \citep{Martin2024MNRAS.535.2375M}. To satisfy these competing goals, we use the Hydrangea suite of cosmological hydrodynamical zoom-in cluster simulations \citep{bahe_hydrangea_2017}.

Hydrangea is based on a DM-only parent simulation \citep{barnes_redshift_2017} of volume (3200 cMpc)$^3$ and DM particle mass of $8 \times 10^{10}\,\mathrm{M}_\odot$. The simulation assumes a flat $\Lambda$ cold dark matter ($\Lambda$CDM) cosmology, with parameters $h \equiv H_0/(100\ \mathrm{km}\, \mathrm{s}^{-1}\, \mathrm{Mpc}^{-1}) = 0.6777$, $\Omega_{\Lambda}$ = 0.693, $\Omega_{M}$ = 0.307, and $\Omega_{b}$ = 0.04825 \citep{planck_collaboration_planck2013_2014}. From this parent simulation, cluster candidates were selected to be re-simulated using a halo mass threshold of $M_{\mathrm{200c}} > 10^{14}$ \,M$_{\odot}$, where $M_{\mathrm{200c}}$ is the mass contained within $R_{\mathrm{200c}}$, the radius inside which the density is 200 times the critical density. An isolation criterion was also applied, such that regions are centred on the local density peak, and not on objects in the proximity of a more massive halo. The final Hydrangea simulations consist of 24 large zoom-in regions each centred on a massive cluster, with $M_{\mathrm{200c}}$ in the range of 10$^{14.0}$ -- 10$^{15.4}$\,M$_{\odot}$. In total, 46 clusters ($M_{\mathrm{200c}} > 10^{14.0}\,\mathrm{M}_\odot$) are found within these 24 regions, which we analyse at redshift $z=0$.

The Hydrangea simulations implement the EAGLE galaxy formation model described in \citet{schaye_eagle_2015}, implemented in a modified version of the \textsc{GADGET}-3 smoothed particle hydrodynamics (SPH) code. Hydrangea adopts a baryonic particle mass of $m_{\rm{baryon}} = 1.81 \times 10^6\,\rm{M}_{\odot}$ ($m_{\rm{DM}} = 9.7 \times 10^6\,\rm{M}_{\odot}$ for DM), and a gravitational softening length of $\epsilon$ = 0.7 physical kpc at $z < 2.8$. Radiative cooling and heating follows the element-by-element prescription of \citet{wiersma2009MNRAS.393...99W}, including a UV/X-ray background \citep{HaardtMadau2001}. Star formation is implemented stochastically, following a pressure law \citep{schaye2008MNRAS.383.1210S} with a metallicity dependent threshold \citep{Schaye2004}. Stellar feedback is modelled through stochastic thermal energy injections \citep{dallavecchia2012MNRAS.426..140D}, while chemical enrichment from Type II/Ia supernovae and AGB stars is followed for 11 elements following \citet{wiersma2009chemical}. Black hole growth and AGN feedback follow \citet{booth2009MNRAS.398...53B}, with Hydrangea adopting the `AGNdT9' model of \citet{schaye_eagle_2015}, in which AGN heat gas stochastically by $\Delta T = 10^9\,\mathrm{K}$.

Haloes and substructures are initially identified using \textsc{Subfind} \citep[see also \citealt{springel_populating_2001}]{dolag_substructures_2009}. First, friends-of-friends (FoF) groups are found by linking DM particles that are spatially close together, within a linking length defined to be 0.2 times the mean inter-particle separation; baryon particles are attached to the FoF group of their nearest DM particle, if any. In a second step, gravitationally self-bound `subhaloes' are identified within each FoF group, based on the detection of local overdensities and an iterative gravitational unbinding procedure, considering both DM and baryon particles. This approach is known to miss the extended haloes of satellites (e.g.~\citealt{Muldrew2011, ForouharMoreno2025}), so we re-assign particles to subhaloes with \textsc{Cantor} (Bahé et al.~in prep.). In contrast to \textsc{Subfind}, this code also considers the past subhalo membership information to define a ``source'' set of particles for each subhalo. As a result, this source set is not limited by the isodensity contour that separates the subhalo from its parent, as is the case for \textsc{Subfind}. Iterative unbinding is then applied to define the final bound subhalo, which therefore includes all particles that are physically part of it, rather than only the dense central core. This approach is similar to the recently developed \textsc{HBT-HERONS} structure finder \citep{ForouharMoreno2025}, but without the need for as many snapshots that prevents us from running \textsc{HBT-HERONS} itself on Hydrangea. Subhaloes contain, in general, both DM and baryon particles (gas, stars, black holes).

\subsection{Identifying the (BCG+)ICL component}
\label{sec: identifying ICL}
Stars in a Hydrangea galaxy cluster can belong to one of three components: the central galaxy (BCG), of which there is exactly one per cluster, one of the (typically numerous) satellite galaxies, or the diffuse halo of ICL. Separating the BCG and ICL components is challenging: their surface brightness profiles are connected smoothly \citep{cui_characterizing_2014, pillepich_halo_2014, montes2022NatAs...6..308M} so that any distinction between them would be physically ambiguous. We therefore do not attempt such a separation and instead consider the combined BCG+ICL, i.e.~the smooth stellar background halo of the cluster. Nevertheless, we can identify approximate (elliptical) radii\footnote{Throughout this work, we use an elliptical or ellipsoidal (in 3D) halo model, in which distances from the centre are expressed through the elliptical radius $r_e \equiv \sqrt{a^2 + b^2\,[+c^2]}$, where $a$, $b$, and (in 3D) $c$ are the projections of a distance vector on the principal axes. For clarity, we refer to this radius with the symbol $a$, i.e.~the length measured along the major axis.}, $a$, at which the BCG or ICL dominate, respectively, for the interpretation of our results. We adopt $a < 0.02\,R_{\mathrm{200c}}$ as a conservative boundary for the BCG region (i.e.~$\lesssim$ 20--40 kpc given that $R_\mathrm{200c} \approx$ 1--2 Mpc for our clusters). This is consistent with \citet[see their fig.~8]{Dolag2010}, who show that the low velocity dispersion component of the BCG+ICL, plausibly associated with the BCG, dominates within this radius. Conversely, we take $a > 0.05\,R_\mathrm{200c}$ ($\gtrsim\,$50--100 kpc) as dominated by the ICL, motivated by the BCG-ICL transition region defined by observations (e.g.~\citealt{Zibetti2005MNRAS.358..949Z, Zhang2019, demaio_growth_2020}).

\begin{figure*}
	\includegraphics[width=2\columnwidth]{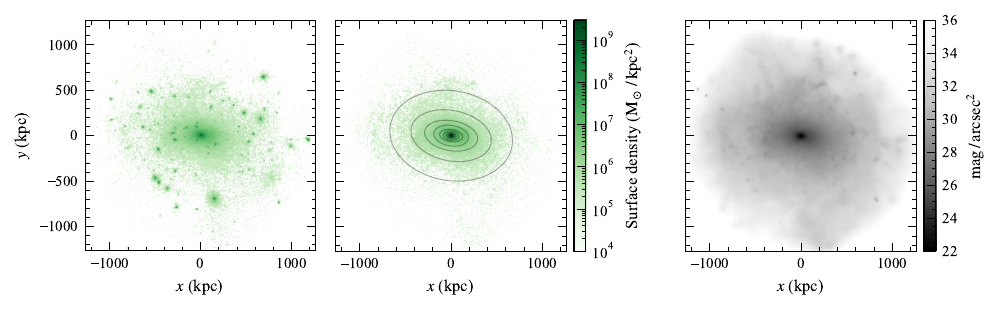}
    \caption{Left: surface density map of the stars in cluster CE-7 G-87 from the Hydrangea simulation at redshift $z = 0$, projected along the $\mathit{xy}$ plane. Middle: surface density map of the BCG+ICL stars, defined by removing stars bound to satellite galaxies, with ellipses indicating the shape measured in several radial bins. Right: surface brightness map in the $H_{\mathrm{E}}$ band of the corresponding \textit{Euclid}-like mock image of the isolated BCG+ICL component (excluding satellites), produced following the methods in \citet{martin_preparing_2022}.}
    \label{fig:ICL extraction}
\end{figure*}

We must, however, separate the BCG+ICL from stars in satellite galaxies. This decomposition is also non-trivial in detail, even in simulations; there is no universally accepted method (see e.g.~\citealt{Pillepich2018, Brown2024MNRAS.534..431B}), primarily due to the diversity of subhalo finders used for different simulations and their subtle differences in what is considered a `satellite'. For Hydrangea, we use the definition of \textsc{Subfind} and \textsc{Cantor} as described above, and consider as the BCG+ICL component all those star particles that are not part of any resolved satellite subhalo. As illustrated in the left-hand and central panels of Fig.~\ref{fig:ICL extraction}, excluding these satellite particles yields a smooth BCG+ICL distribution without obvious clumps or holes. The same is true for mock surface brightness maps (right-hand panel, generated as described in Section \ref{sec:isophotal_fitting}), albeit with slightly more pronounced substructure due to the inhomogeneous distribution of young bright stars. An analogous decomposition is applied to the DM component: all particles associated with satellite subhaloes are removed, leaving the smooth ``main halo'', which we hereafter refer to simply as DM. We note that this approach is fundamentally different from observational analyses, where satellites must be masked based on photometric information rather than on whether or not they are bound to a satellite. In Section\,\ref{sec:sim_comparison}, we explore how different methods to mask stars belonging to the satellites can impact the measured shape of the ICL and DM halo.

\subsection{Defining the shape of the DM and BCG+ICL haloes}
\label{defining_BCG+ICL}
We quantify the shape of the BCG+ICL and DM using the mass-weighted moment-of-inertia tensor, otherwise known as the shape tensor, which characterises how mass is distributed around a central point. This approach is commonly used to determine the shapes of galaxies and DM haloes in simulations (e.g.~\citealt{zemp_determining_2011,chua_shape_2019, pulsoni_stellar_2020,valenzuela_galaxy_2024}). Apart from its relative simplicity, one key appeal of the shape tensor is that it uses particle data directly and thus avoids the subjective choices often needed in model-fitting approaches, such as the selection of brightness thresholds assumed in functional forms for the light distribution. We use the simple unweighted shape tensor here, rather than its reduced version that weights particles by their inverse radius; our analysis is performed within thin ellipsoidal shells, so any differences between the two are expected to be negligible. The shape tensor is therefore defined as

\begin{equation}\label{eqn:tensor element}
    S_\mathit{ij} = \frac{\sum{(M_n\, x_\mathit{n,i}\, x_\mathit{n,j}})}{\sum M_n},
\end{equation}

where $i$ and $j$ denote the spatial axes, $x_\mathit{n,i}$ is the coordinate of the $n^{\mathrm{th}}$ particle in the shell relative to the cluster centre\footnote{Throughout, we define the cluster centre as the coordinates of the particle with the lowest gravitational potential.} along axis $i$, and $M_n$ its mass. Diagonalising $S_\mathit{ij}$ returns its eigenvectors, which correspond to the principal axes of the ellipsoid. The corresponding eigenvalues are proportional to the squared axis lengths ($a$, $b$, and $c$) corresponding to the major, intermediate, and minor axes, respectively ($a > b > c$). From these, the axis ratios $p = b/a $ and $q = c/a$ are computed. In this notation, $p=q=1$ corresponds to a perfect sphere, while smaller values of $p$ and/or $q$ represent progressively more elongated ellipsoids. These two ratios, along with the directions of the major and minor axes\footnote{As the intermediate axis is perpendicular to both the major and minor axes, its direction is unambiguously specified by these two.}, form the basis of our shape analysis.

We apply an analogous method in two dimensions. The only difference is that only two axes are involved here (major and minor), resulting in a single axis ratio ($p = b/a$). Likewise, there is now only one relevant orientation -- that of the major axis -- which can be conveniently described by its position angle $\phi$.

Since the shape of the system is not known a priori, an iterative process is necessary to select the particles included in each shell. We begin with a spherical shell with a given inner radius $r$ and thickness $\Delta r$, and compute the shape tensor for the particles contained in this volume. We then derive the axis ratios and directions as detailed above, and use these to define a new, ellipsoidal, shell with semi-major axis $a = r$ and width $\Delta a = \Delta r$. The process is then repeated for the particles contained in this shell, iterating until the relative change in successive values of $p$ and $q$ is less than 0.005. We divide the halo into 30 logarithmically spaced bins from 3\,kpc to $R_{\mathrm{200c}}$. In each, we measure, independently, the axis ratios and orientations of the BCG+ICL and DM, as described above.

To test the accuracy of this method and inform our choice of $r$ and $\Delta r$, we have applied it to sets of mock particles generated according to ellipsoidal distributions with known axis ratios and orientations. Points are sampled randomly-uniformly in a sphere and then transformed via an analytic mapping to follow a target power-law density profile $\rho (r) = \alpha r^{-\beta}$, where $\alpha$ and $\beta$ are constants that describe the normalisation and slope, respectively. The $y$ and $z$ coordinates of the particles are then scaled according to the target values of $p$ and $q$, respectively, before rotating them in the desired (arbitrary) orientation. The two quantities that primarily affect the accuracy are the number of particles in each shell, $N_{\rm shell}$ (which depends on $\alpha$ and $\beta$ for fixed values of $r$ and $\Delta r$) and, perhaps less obviously, the slope of the density profile, $\beta$. The latter is important, even at fixed $N_\mathrm{shell}$, because a steeper profile corresponds to a more pronounced elliptical shape\footnote{In the extreme case of a uniform density profile, there is no meaningful shape at all.} and hence allows a more accurate determination of the axis ratio and orientation.

For our test, we set up a 3D particle distribution representing an ellipsoid with an arbitrarily chosen semi-major axis length of 1000\,kpc with axis ratios $p=0.7$ and $q=0.5$, and a rotation given by Euler angles $(\Phi, \Theta, \Psi) = (30^\circ, 30^\circ, 30^\circ)$. We consider multiple values for $N_{\rm shell}$ and $\beta$ over the range encountered in Hydrangea ($10^{3.7}<N_{\rm shell}<10^{5.7}$ and $1.0<\beta<3.5$). For each combination, we generate a suite of 50 identical models that differ only in the seed of the random number generator, and compute the axis ratio and axis orientation with our iterative shape tensor method within a shell of $r = 250$\,kpc and width $\Delta r = 50$\,kpc. We then compute the fractional root-mean-square deviation of the recovered values from the true input as an estimate of our measurement uncertainty. At $0.1\,R_{\mathrm{200c}}$, the median values of $N_{\rm shell}$ and $\beta$ for the ICL in Hydrangea are $3.81 \times 10^{4}$ and 2.87, respectively, while for the DM they are $2.76 \times 10^{5}$ and 1.48. These correspond to uncertainties of approximately 0.75 per cent on $p$ and 0.79 per cent on $q$ for the ICL, with position angle uncertainties of $\approx$0.6 degrees. For DM, the typical uncertainties are 1.2 per cent ($p$), 1.9 per cent ($q$), and $\approx$0.4 degrees for the position angles. An analogous process is used in 2D, yielding uncertainties that are comparable to their 3D analogues. We also test the impact of plausible offsets between the assumed reference centre of the shape measurement method and the true mass distribution using mock ellipsoids, and find no significant effect on the recovered shapes in the ICL-dominated region ($a \gtrsim 0.05\,R_\mathrm{200c}$).
The full uncertainty distributions and details of the offset tests are provided in Appendix \ref{app: error grids}.

\section{Does the ICL trace the DM halo shape?}
\label{sec: ICL DM shapes}

\subsection{The 3D shape of BCG+ICL compared to DM}
\label{sec:3D}
\begin{figure*}
	\includegraphics[width=\textwidth]{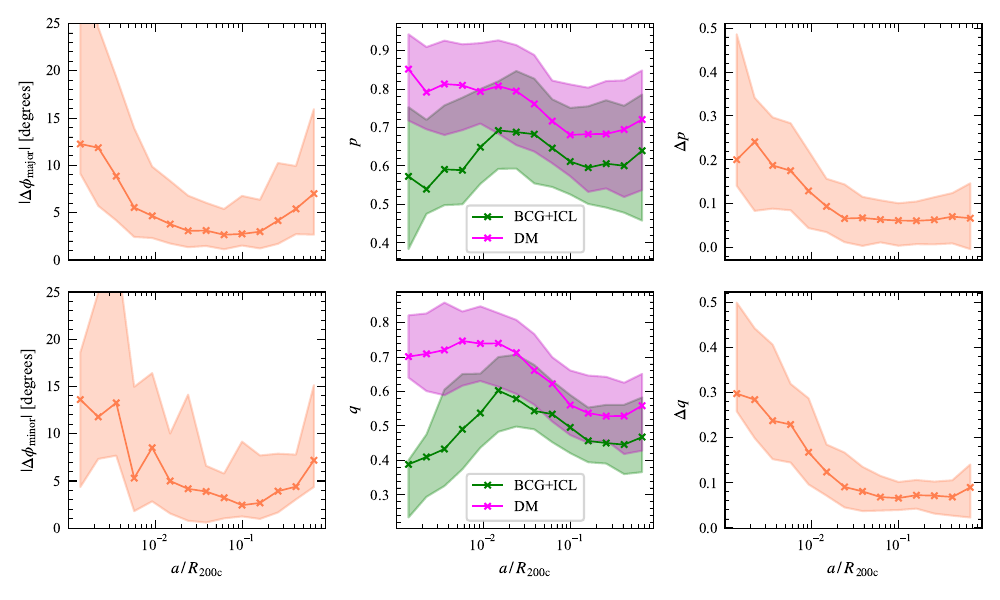}
    \vspace{-0.3cm}
    \caption{Radial profiles of 3D axis orientations and ratios for BCG+ICL and DM. Left column: (absolute) angle between the BCG+ICL and DM major axes, $|\Delta \phi_{\mathrm{major}}|$ (top), and minor axes, $|\Delta \phi_{\mathrm{minor}}|$ (bottom), respectively. Middle column: median axis ratios $p = b/a$ (major to intermediate; top), and $q = c/a$ (major to minor; bottom) for the BCG+ICL (green) and DM (purple). Right column: median difference in axis ratios $p$ (top) and $q$ (bottom) between the BCG+ICL and DM. All three quantities are plotted against the normalized (elliptical) radius, $a/R_{\mathrm{200c}}$. Solid lines trace the medians, shaded regions indicate the 1$\sigma$ scatter. Both the orientation and axis ratio agree closely between the BCG+ICL and DM, albeit with a slightly more elliptical shape for the former.}
    \label{fig:3D results}
\end{figure*}

With our measurements of axis orientation and ratios in place, we first examine the 3D shapes of galaxy clusters, as they provide a direct view of the physical BCG+ICL and DM components without complications from line-of-sight projection effects. In the top-left panel of Fig.~\ref{fig:3D results}, we show the radial profile of the angle $\Delta \phi_\mathrm{major}$ between the major axes of the BCG+ICL and DM distributions, i.e.~the degree of misalignment between them. The individual measurements from the 30 bins of each clusters are re-binned to a common grid of 14 points spaced logarithmically between $10^{-3}$ and $1$ in $a / R_\mathrm{200c}$; we then show the median and 1$\sigma$ scatter of all these profiles. The bottom-left panel shows the analogous difference $\Delta \phi_\mathrm{minor}$ between the minor axes of the BCG+ICL and DM.

Both angles are typically small, with a minimum in the median of only a few degrees around $0.1\,R_\mathrm{200c}$. The misalignment angles increase slightly towards smaller and larger radii, reaching a maximum of $\approx$13 degrees in the centre-most bin and $\approx$8 degrees at $R_\mathrm{200c}$. The scatter increases likewise, but the 1$\sigma$ upper bound remains below $\approx$15 degrees in the ICL region. The less tight alignment in the centre, deep in the BCG, is plausibly the result of galaxy formation physics, whereas the slight difference near the outer edge may plausibly reflect deviations from a pure ellipsoidal shape due to recently accreted satellites. Overall, however, the orientations of both the major and minor axes agree closely between the BCG+ICL and DM.

Based on this close agreement in the axis orientations, it is meaningful to investigate the axis ratios -- if the two components were significantly misaligned, one would not expect a close correspondence in the latter. In the middle column of Fig.~\ref{fig:3D results}, we display the radial profiles of axis ratio $p$ (top) and $q$ (bottom), for the BCG+ICL (green) and DM (purple). Profiles from individual clusters are stacked in analogy to the misalignment angle described above. The two profiles are close, and overlap within their 1$\sigma$ scatters over most of the radial range probed here, but they are not identical: both the major-to-intermediate ($p$) and the major-to-minor axis ratio ($q$) are systematically smaller for the BCG+ICL than for the DM, implying that the former is more elongated than the latter. The BCG+ICL axis ratio profiles are also non-monotonic, being roundest near the BCG-to-ICL transition ($\approx 0.02\, R_{\mathrm{200c}}$), slightly more elongated at larger radii, and significantly more so towards the centre of the BCG. In contrast, the DM is roundest in the centre and then becomes consistently more elliptical towards the cluster outskirts. As a result, the two profiles are closest to each other in the ICL domain ($a \gtrsim 0.05\,R_\mathrm{200c}$) and most discrepant near the centre.

To demonstrate the difference in the axis ratios more clearly, the right column of Fig.~\ref{fig:3D results} shows the radial profiles of the difference between the BCG+ICL and DM, $\Delta p = p_{\rm{DM}} - p_{\rm{BCG+ICL}}$ (top) and $\Delta q = q_{\rm{DM}} - q_{\rm{BCG+ICL}}$, (bottom), respectively, measured for each cluster individually\footnote{This is therefore not, in detail, the same as the difference between the purple and green curves in the middle column, which represent the median over all clusters.}. Confirming the conclusion from the middle panels, these offsets are greatest near the centre (medians of $\approx$+0.2 and +0.3 for $\Delta p$ and $\Delta q$, respectively, i.e.~with the DM being appreciably rounder), accompanied by substantial scatter in both. Combined with the increased misalignment in the central region, this indicates that the BCG is not a reliable tracer of the innermost shape of the DM halo, at least in Hydrangea. As mentioned above, this discrepancy is plausibly the result of galaxy formation physics and may therefore reflect limitations of the simulation model (see Section\,\ref{sec:sim_comparison}) -- as shown by \citet{bahe_hydrangea_2017}, the stellar mass of BCGs in Hydrangea is a factor $\approx$2--3 higher than inferred from observations. In the ICL-dominated region, however, both $\Delta p$ and $\Delta q$ decline to roughly constant values of $\approx 0.07$, with scatter of less than 0.1. At $R_{\mathrm{200c}}$, only 8 out of the 46 clusters have DM components that are slightly more elongated than their ICL counterparts (i.e. $\Delta p < 0$), although the offsets remain small and close to zero. This small systematic offset indicates that the ICL and DM axis ratios are in quite close agreement, but the DM is nevertheless slightly rounder. Together with the small offset in the axis orientations in the ICL regime, these results therefore imply that the ICL can trace the intrinsic DM halo shape.

\subsection{Comparison of shapes in 2D}
Having established a relatively close agreement between the BCG+ICL and DM shapes in three dimensions, we now proceed to investigate how this similarity translates to shapes measured in 2D projection, since we are only able to measure those in observations. Our goal here is not to make a detailed mock observational analysis (though see Section \ref{sec:isophotal_fitting} below), but simply test whether projected BCG+ICL measurements can, in principle, be used as a proxy for the DM shape. As such, we still use the particle coordinates but project them, for each cluster, along three orthogonal axes (onto the $xy$, $xz$ and $yz$ planes). Shapes are then measured in analogy to the 3D case, as described in Section \ref{defining_BCG+ICL}.

\begin{figure*}
	\includegraphics[width=\textwidth]{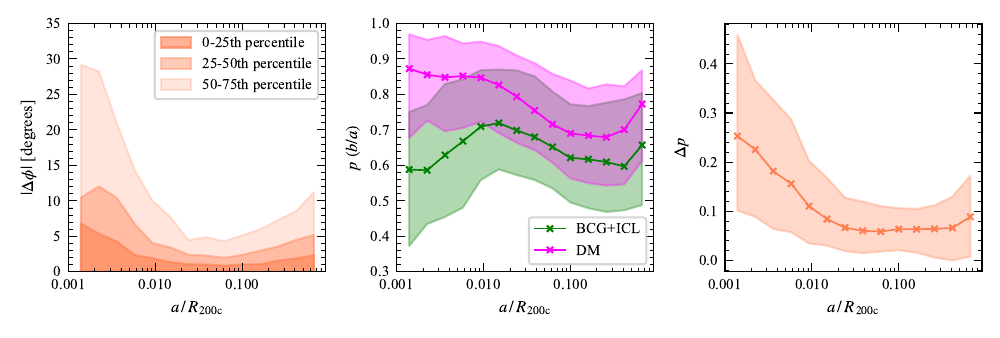}
    \vspace{-0.3cm}
    \caption{Radial profiles of orientations and axis ratios of BCG+ICL vs.~DM, analogous to Fig.\,\ref{fig:3D results} but in 2D projection. Left: absolute angle difference, $|\Delta \phi|$, between the position angles of the DM and ICL major axes against $a/R_{\mathrm{200c}}$. Shaded regions represent the 0--25$^\mathrm{th}$, 25$^\mathrm{th}$--50$^\mathrm{th}$, and 50$^\mathrm{th}$--75$^\mathrm{th}$ percentiles. Middle: major-to-minor axis ratio, $p=b/a$, of the BCG+ICL ($p_{\rm{BCG+ICL}}$, green) and DM ($p_{\rm{DM}}$, magenta). Solid lines trace the medians, shaded regions the 1$\sigma$ scatter. Right: axis ratio difference $\Delta p = p_{\rm{DM}} - p_{\rm{BCG+ICL}}$. As in 3D, position angles and axis ratios agree closely, with the tightest correspondence at ICL-dominated radii.}
    \label{fig:diff plot Hydrangea}
\end{figure*}

The result is shown in Fig.~\ref{fig:diff plot Hydrangea}, whose layout is similar to Fig.~\ref{fig:3D results} but only has a single row because there is only one position angle and axis ratio in this case (see above). The left-hand panel shows the difference in the position angle of the BCG+ICL and DM major axes, as a function of normalised radius $a/R_{\mathrm{200c}}$. In contrast to the 3D case, the absolute position angle is now defined by a single number so that the difference can be positive or negative and the median will, by symmetry, be close to zero. We are interested, instead, in the absolute deviation from zero, $|\Delta \phi|$, which is what is plotted. However, because the full distribution (positive and negative) of $\Delta \phi$ peaks around zero, the same is true for $|\Delta \phi|$, so that the latter has a highly non-Gaussian shape. The median and 1$\sigma$ scatter, as plotted for the 3D case in Fig.~\ref{fig:3D results}, are therefore not a particularly meaningful representation of this distribution. We show instead the first, second, and third quartiles of $|\Delta \phi|$, in increasingly fainter shades of orange; the median absolute deviation therefore corresponds to the upper end of the mid-orange band. The result is very similar to what we had found in our 3D analysis above: the tightest alignment is seen at intermediate radii (a median absolute difference of $\approx$ 2--3 degrees, here at $\approx$0.05 $R_\mathrm{200c}$ which is slightly smaller than for 3D), with a mild increase to the outskirts ($\approx$5 degrees) and a stronger discrepancy in the centre (up to 17 degrees). These values are very close to their 3D counterparts, indicating that projection effects alone have a negligible influence on the measured (mis-)alignment between the BCG+ICL and DM. The approximately even spacing between the three bands indicates significant cluster-to-cluster scatter, with some aligned near-perfectly even in the centre while 25 per cent deviate by more than 30 degrees. Overall, however, we see a close alignment between the two components, especially in the ICL-dominated region.
 
The middle panel of Fig.~\ref{fig:diff plot Hydrangea} presents the median projected axis ratios, $p=b/a$, of the BCG+ICL and DM components. The shape of each component mirrors their 3D analogue, with the possible exception of a slightly more pronounced upturn in the outermost bin (in both the BCG+ICL and DM). Quantitatively, the axis ratios are somewhat closer to the major-to-intermediate axis ratio in 3D ($p$; top panel in Fig.~\ref{fig:3D results}) than to the major-to-minor one ($q$). As in the 3D analysis, the DM is systematically rounder, with a higher median axis ratio at all radii but particularly near the centre. Nevertheless, the 1$\sigma$ scatter regions overlap over the full radial range that we probe. The corresponding difference in axis ratios, $\Delta p = p_{\rm{DM}} - p_{\rm{BCG+ICL}}$, is shown in the right-hand panel. As for the 3D case, $\Delta p$ is highest and has the largest scatter in the very centre (a median difference of 0.25, with 1$\sigma$ range from 0.10 to 0.46). The difference then decreases with increasing distance from the cluster centre and reaches a plateau of $\approx$0.07 (median) in the ICL-dominated region, with a much smaller scatter of only $\approx$0.056. This behaviour is, again, quantitatively consistent with the trends seen in 3D (Fig.~\ref{fig:3D results}, right column), confirming the potential of tracing the DM halo shape through observations of ICL in projection. For the rest of this work, we focus on the 2D measurements since these are more directly comparable to observations.

We also measure the shapes for a subset of projected clusters using all DM particles in the halo, rather than the main halo DM defined in Section \ref{sec: identifying ICL}. We find that this does not significantly alter the overall trends in axis ratios or position angles. For consistency and to focus on the relaxed halo component, we therefore retain the main halo DM for all subsequent analysis.

\subsection{Sensitivity to mass and formation history}

The shapes of DM cluster haloes are known to depend on properties such as mass (e.g.~\citealt{KasunEvrard2005}) and formation redshift (e.g.~\citealt{despali_like_2014}). This is because clusters grow through mergers, redistributing the existing BCG+ICL and DM material while simultaneously adding new mass from the infalling system(s). It is therefore of interest to investigate whether the tight correlation between ICL shape and DM halo shape holds for clusters of all masses and assembly stages. To do this, we divide the Hydrangea cluster sample into a low-mass ($M_{200} < M_{\rm median}$) and high-mass ($M_{200} \geq M_{\rm median}$) sample, where $M_{\rm median} = 1.84 \times 10^{14}\,\mathrm{M}_\odot$ is the median mass of our sample. We similarly divide the sample into low- and high-$z_{\mathrm{50}}$ subsets, where $z_{\mathrm{50}}$ is the redshift at which the cluster assembled half of its present-day halo mass.

\begin{figure*}
	\includegraphics[width=2\columnwidth]{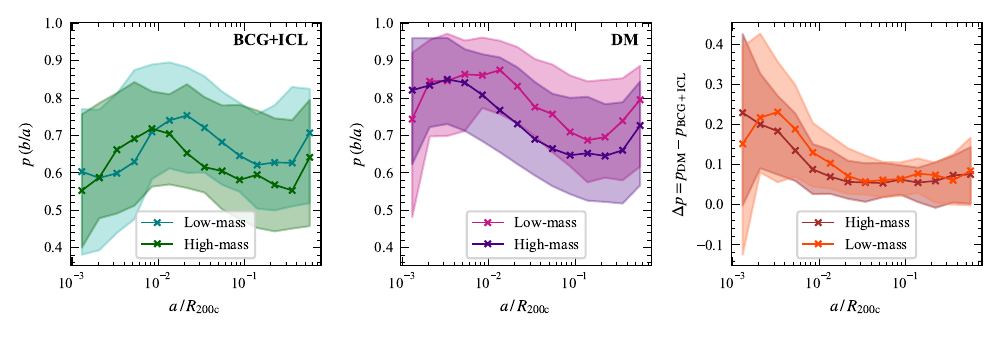}
    \caption{Radial profiles of projected axis ratios in low- and high-mass clusters. Left: axis ratios $p$ of the BCG+ICL, as a function of $a/R_{\mathrm{200c}}$, for Hydrangea clusters below (blue) and above (green) the median mass of the full sample ($1.8\times 10^{14}\,\mathrm{M}_\odot$). Middle: the same, but for DM (pink and purple for low- and high-mass clusters, respectively). Right: difference between the BCG+ICL and DM axis ratios for high- (brown) and low-mass (orange) clusters. In all three panels, solid lines represent medians and shaded regions the 1$\sigma$ scatter. The offset between BCG+ICL and DM is very similar in both subsamples, despite clear trends in either component individually with halo mass.}
    \label{fig:Mass split}
\end{figure*}

The result, for the split by halo mass, is shown in Fig.~\ref{fig:Mass split}. As expected from previous works (e.g.~\citealt{KasunEvrard2005,despali_like_2014, lau_correlations_2020}), the axis ratios of the low- and high-mass subsamples are systematically offset from each other with the low-mass clusters generally being rounder in both the BCG+ICL (left) and DM (middle), especially at $a \gtrsim 0.01\,R_\mathrm{200c}$. However, the offset is small ($\lesssim$0.1) and the 1$\sigma$ scatter regions of both subsamples overlap substantially. More important for our purpose, the right panel displays the corresponding differences in axis ratio between BCG+ICL and DM, $\Delta p$, for the low- and high-mass subsets. Both trace each other very closely, especially in the ICL region where the medians and scatter differ by less than 0.01. At $a \lesssim 0.02\,R_\mathrm{200c}$, i.e.~in the BCG-dominated region, there is a slight systematic offset in the sense that $\Delta p$ is higher (by up to $\approx$0.06) for low-mass clusters -- in other words, their BCG+ICL is slightly more elongated than the DM, compared to their massive counterparts. Inspecting the left and middle panels of Fig.~\ref{fig:Mass split}, this is most likely due to a slightly stronger decline in $p$ for the stellar component in low-mass clusters, plausibly due to the higher susceptibility to baryon effects such as AGN feedback. Overall, however, in the ICL-dominated region that is the focus of this work, we find no dependence of $\Delta p$ on halo mass, indicating that the ability of the ICL to trace the DM is not affected by halo mass.

We have also compared the corresponding misalignment angles ($|\Delta \phi|$) for the two subsets, which are not shown in Fig.~\ref{fig:Mass split}. Beyond $\approx 0.01\,R_\mathrm{200c}$ both show excellent alignment ($|\Delta \phi| < 5$ degrees), with a slightly smaller offset for the massive clusters.

We have performed a similar comparison splitting the clusters by formation redshift. Here, we find no significant difference in the axis ratios $p$ of the BCG+ICL and DM components, nor in $\Delta p$, between the high- and low-$z_{\mathrm{50}}$ subsamples (not shown). This contrasts with previous works (e.g.~\citealt{despali_like_2014}) that have found higher ellipticities (i.e.~lower axis ratios) for more recently assembled haloes, although this discrepancy can plausibly be attributed to our small sample of extremely massive haloes as well as the simplified way in which $z_{\mathrm{50}}$ reflects a cluster's formation history. Even if a more careful comparison with a significantly larger sample may recover the previously reported (small) offset in absolute axis ratio, the main conclusion for our purpose is robust: neither cluster mass nor formation redshift significantly affects the agreement in axis ratio between the BCG+ICL and DM.

To probe the recent accretion history of the clusters, we split our sample into four bins according to $M_{12}$, defined as the stellar mass ratio between the BCG and the most massive satellite galaxy in the cluster \citep{kimmig_2025}. This quantity acts as a proxy for the merger and recent accretion history of the cluster. We find that the clusters in the bin with the highest $M_{12}$ values exhibit slightly more elliptical BCG+ICL and DM at $a \geq 0.2$\,$R_{\mathrm{200c}}$, likely corresponding to the regions where the most massive satellites tend to reside. However we find no significant trend in axis ratio and position angle differences across the $M_{12}$ bins. We therefore conclude that the recent accretion history, as traced by $M_{12}$, does not affect the reliability of the ICL as a tracer of the underlying DM shape.

\subsection{Measuring the shape in mock ICL images}
\label{sec:isophotal_fitting}
We have so far relied on the shape tensor method applied to simulation particles to quantify the orientation and axis ratio of our haloes. This approach is not feasible for  observations, where the shape of the ICL is commonly determined via isophotal fitting (e.g.~\citealt{kluge_structure_2020, Montes2021, kluge_euclid_2025}). To test how robust our conclusion about the agreement between BCG+ICL and DM shapes are to this difference in methodology, we repeat our (2D) analysis but with the BCG+ICL shapes determined by fitting ellipses to isophotal contours in mock images.

To this end, we produce mock images of the BCG+ICL for each Hydrangea cluster and projection, assuming a redshift\footnote{This redshift is broadly representative of early ICL studies with \textit{Euclid} \citep{bellhouse2025A&A...698A..14E} but its precise value is unimportant here.} of $z = 0.2$, following the methods described in \cite{martin_preparing_2022}. Since \textit{Euclid} will soon enable isophotal ellipticity measurements of thousands of clusters \citep{bellhouse2025A&A...698A..14E}, we match our mock images to the \textit{Euclid} image specifications. While the measured shapes are consistent across all \textit{Euclid} bands, the ICL can be traced to the largest radii from the cluster centres in $H_{\mathrm{E}}$ images (Kolcu et al., in prep.). We therefore adopt this band for all images used in this work. The BCG+ICL stellar particles are extracted as detailed in Section \ref{defining_BCG+ICL}. For each particle with $r < R_\mathrm{200c}$, spectral energy distributions (SEDs) are computed based on its age and metallicity using the models of \cite{bruzual_stellar_2003}. Luminosities are derived by redshifting and convolving the SEDs with the \textit{Euclid} $H_{\rm{E}}$ band transmission function to produce the apparent magnitudes. To suppress artificial pixel-to-pixel fluctuations in low-density regions, the luminosity is adaptively smoothed with a Gaussian kernel whose standard deviation is equal to the distance to the fifth-nearest neighbour (star) particle. For each cluster, three images are produced in the $xy$, $xz$, and $yz$ projections and convolved with a Gaussian kernel with FWHM = 0.49\,arcsec, consistent with the extended $H_{\rm{E}}$ band \textit{Euclid} point spread function (PSF) measured from the Early Release Observations \citep{Cuillandre2025}. For simplicity and generality, we do not add noise or observational artefacts. Finally, the image is binned to a spatial resolution of 0.3 arcsec pixel$^{-1}$, as for \textit{Euclid}. As demonstrated in Fig.~\ref{fig:ICL extraction}, the resulting mock images are qualitatively similar to the mass surface density maps, albeit with somewhat more pronounced substructure due to the presence of young, bright stars.

We derive radial shape profiles from the mock images of the BCG+ICL using the light-profile fitting software \texttt{AUTOPROF} \citep{stone_autoprof_2021}, which fits a sequence of elliptical isophotes to the light distribution of the cluster. The code is provided with fixed central coordinates, which we take as those of the most bound particle in each cluster. The background level is fixed to zero, and the semi-major axis of the first isophote is set to one pixel, growing by 0.15 dex for each subsequent ellipse.

\texttt{AUTOPROF} first computes global estimates of the position angle and axis ratio, $b/a$, from a Fourier decomposition of the light within circular apertures. Using these initial values, the code then generates a series of concentric ellipses and iteratively adjusts their position angles and ellipticities through a regularised optimisation scheme. In each iteration, a loss function is minimised that combines the regularisation term with the power in the second Fourier mode, which measures how well an ellipse follows the local light distribution. Upon convergence, this method yields a stable set of isophotes that provide radial profiles of the position angle $\phi$ and axis ratio, out to the (elliptical) radius where the per-pixel surface brightness reaches the $1\sigma$ background noise estimated internally by \texttt{AUTOPROF}. Beyond this limit the shape parameters are kept fixed at their last reliable values, and the isophote sequence is continued to the edge of the image. We limit our comparison for each cluster to radii where \texttt{AUTOPROF} can measure the axis ratio and position angle, however these extend out to $R_{\rm 200c}$ for all clusters.

\begin{figure}
        \includegraphics[width=1\columnwidth]{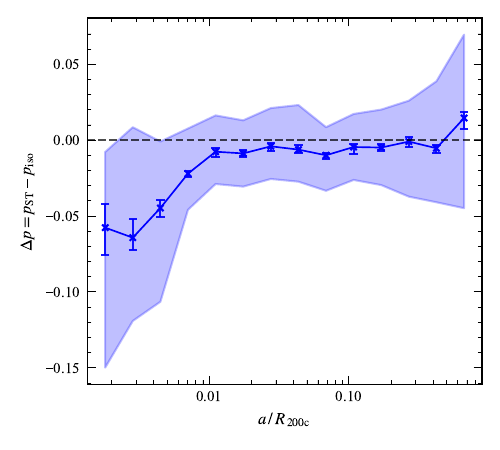}
    \caption{Difference in BCG+ICL axis ratio derived from the shape tensor of star particles, $p_{\rm ST}$, and from isophote fitting of mock images with \texttt{AUTOPROF} applied to corresponding mock images, $p_{\rm{iso}}$. The solid line shows the median, the shaded region the 1$\sigma$ scatter, and the error bars show 1$\sigma$ uncertainties estimated via bootstrap resampling. Beyond the very centre ($a \gtrsim 0.01\,R_\mathrm{200c}$), both methods give near-identical results.}
    \label{fig:isophotal_results}
\end{figure}

The absolute difference in the BCG+ICL position angle measured with \texttt{AUTOPROF} and our default shape tensor approach (in 2D) is consistently small, with a median of only 2.5 degrees. The orientation of the BCG+ICL, and by extension that of the DM, can therefore be measured reliably with methods accessible to observational data. The same is not quite true for the axis ratios, which are compared in Fig.~\ref{fig:isophotal_results}. The error bars on the median are estimated using bootstrapping, generating 1000 resampled cluster profiles, with the 1$\sigma$ scatter amongst these taken as the uncertainty. Beyond $\approx 0.01\,R_\mathrm{200c}$ the difference between the axis ratios is very small ($\lesssim$0.01), generally in the sense that the shape tensor gives slightly lower (less round) axis ratios. We can therefore conclude that isophotal fitting provides a reliable measurement of the projected shape of the ICL. In the centre, however, the \texttt{AUTOPROF} axis ratios are systematically larger (i.e.~more circular), by up to 0.055. While this is still a minor offset compared to the $\approx$0.2 discrepancy between the BCG+ICL and DM in the same region, precision measurements of the BCG shape must take this method discrepancy into account. Given that the difference is limited to radii not much larger than the assumed PSF and pixel scale (see above), it plausibly originates from these two effects.

\section{Shape dependence on simulation models}
\label{sec:sim_comparison}

\subsection{Impact of satellite removal methods on the shapes of ICL and DM}
\label{sec:sat_removal_comparison}

Since there is no single, universally accepted method for separating the ICL from the satellites, it is important to asses the sensitivity of our results to this choice. We therefore compare the shapes computed using our fiducial particle-level classification to those obtained from a simplified radial exclusion similar to that used by \citet{butler_intracluster_2025}.

In this approach, we use the subhalo catalogue, \textsc{Cantor}, solely to identify satellite galaxies, which we define as subhaloes with a stellar mass $M_\mathrm{star} \geq 10^8\,\mathrm{M}_\odot$. We then excise particles `belonging' to each satellite based on a simple radial aperture $r_\mathrm{sat}$. We take the radius of this aperture as a multiple, $\gamma$, of the stellar half-mass radius $r_\mathrm{50}$ for massive galaxies ($M_\mathrm{star} > 10^{11}\,\mathrm{M}_\odot$), but use a constant value $r_\mathrm{low}$ for less massive ones. The value of $\gamma$ is to some extent arbitrary; we here use $\gamma = 4$ as in \citet{butler_intracluster_2025} but have verified that the exact choice does not strongly impact our results, except in the innermost radial bins where higher values lead to stronger over-subtraction of particles due to the much higher stellar density. The use of a constant value $r_\mathrm{low}$ at low masses likewise follows \citet{butler_intracluster_2025}, who introduce it due to the significantly overpredicted sizes of Horizon-AGN galaxies in this mass range. This value corresponds to the 84$^\mathrm{th}$ percentile of stellar half-mass radii of galaxies with stellar masses between $10^8$ and $10^{11}\,\mathrm{M}_\odot$, $r_\mathrm{low} = 23.6$ kpc. DM and star particles are removed within the same apertures, and the remaining particles classified as the BCG+ICL and DM main halo, respectively. For clarity, we explicitly refer to this simplified definition with the suffix `$4 r_\mathrm{sat}$'. It is important to keep in mind that this method will leave spherical voids and shell-like residuals around each satellite in the BCG+ICL distribution (and also in the DM, albeit to a lesser extent as the DM substructure fraction is much lower). To mitigate the impact of these artefacts, we therefore limit this analysis to 2D, where projection of fore-/background particles partly smoothes them out.

\begin{figure}
    \centering
    \includegraphics[width=1\columnwidth]{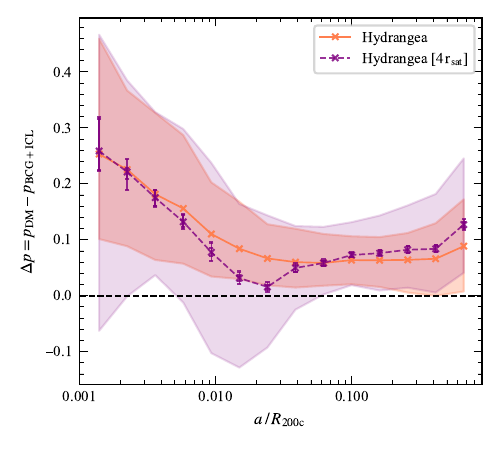}
    \caption{Impact of the satellite removal method on the axis ratio difference between BCG+ICL and DM. The solid orange line shows our fiducial method based on particle-level subhalo assignment, identical to the right-hand panel of Fig.~\ref{fig:diff plot Hydrangea}. The simpler method of excising spheres with radius $4 r_\mathrm{sat}$ around each resolved satellite is represented by the dashed purple line (see text for details). Shaded bands give the corresponding $1\sigma$ scatter, while error bars show 1$\sigma$ uncertainties on the median estimated via bootstrap resampling. The black dashed line indicates a difference of zero, i.e.~perfect agreement between DM and BCG+ICL. Despite qualitative agreement on a small but typically slightly positive offset between the two components, the magnitude of the discrepancy depends on how exactly satellites are removed.}
    \label{fig:Hydrangea Re Method}
\end{figure}

Fig.~\ref{fig:Hydrangea Re Method} demonstrates how this alternative method affects the measured (2D) axis ratio difference, $\Delta p$, for Hydrangea where we have the `true' reference decomposition from \textsc{Cantor}. The error bars on the median are estimated using bootstrapping, following the same method as in Sect.~\ref{sec:isophotal_fitting}. The two excision methods agree very closely in the very centre ($a \lesssim 0.004$\,$R_{\mathrm{200c}}$), but at larger radii there is a slight difference. Notably, the offset is not monotonic: at intermediate radii (out to $\approx0.04\,R_\mathrm{200c}$) the simple spherical excision yields a lower $\Delta p$, by up to $\approx0.05$ in the median and $\approx0.16$ in the 1$\sigma$ scatter, while at larger radii this trend reverses and it results in $\Delta p$ that is higher by a similar amount. The measured systematic offset between the DM and BCG+ICL is therefore, in detail, dependent on the technique used to remove satellites.

\subsection{Effects of formation physics on the shapes of ICL and DM}
\label{sec: simulation_comparison}

Another factor that influences the shapes of the BCG+ICL and the DM is the choice of cosmological simulation and its galaxy formation model. To assess how sensitive the results presented in Section \ref{sec: ICL DM shapes} are to the galaxy formation model of Hydrangea, we repeat our analysis on the Horizon-AGN simulation \citep{dubois_dancing_2014}, which differs from Hydrangea in several key aspects.

Like Hydrangea, Horizon-AGN is a cosmological hydrodynamical simulation, but instead of zooming in on clusters it models a complete periodic volume of (142\,cMpc)$^3$. The simulation is performed with the RAMSES code and contains 1024$^3$ DM particles with a mass resolution of $m_{\rm{DM}} = 8 \times 10^7 \, \rm{M}_{\odot}$ (a factor $\approx$8 higher than in Hydrangea). In contrast to the SPH approach of Hydrangea, RAMSES models the gas on an adaptively refined grid, with a minimum spatial resolution of 1\,kpc. Stars are formed in sufficiently dense regions, based on a local Schmidt law; star particles have an initial mass of $\approx$2$\times 10^6\,\mathrm{M}_\odot$, similar to Hydrangea. A $\Lambda$CDM cosmology is adopted, with $H_0 = 70.4\ \rm{km\ s}^{-1}\ Mpc^{-1}$, $\Omega_{\rm{m}} = 0.272$, and $\Omega_{\rm{b}} = 0.045$; these parameters are based on the \textit{WMAP7} cosmology \citep{komatsu_seven-year_2011} but the slight ($\lesssim$10 per cent) difference from Hydrangea is insignificant for this work. A comprehensive suite of subgrid models covers gas cooling and heating, star formation, and stellar and AGN feedback, as described in \cite{Kaviraj_2017}. The assumptions behind these models, and their implementation details, differ significantly from Hydrangea, allowing an independent test of our previous results. Previous works have shown that Horizon-AGN broadly reproduces bulk properties of galaxies in the real Universe across a wide redshift range ($0<z<6$) without being explicitly calibrated on these \citep{Kaviraj_2017}. This simulation has been extensively used for ICL studies \citep{Canas_2020, brough_preparing_2024, Brown2024MNRAS.534..431B, butler_intracluster_2025, kimmig_2025}, which have demonstrated good overall agreement of its ICL properties with several other simulations. We use the same 12 clusters\footnote{Although there are 14 haloes with $M_{\rm 178c}>10^{14}\,{\rm M}_\odot$ at $z=0.06$, the two most massive clusters are undergoing major mergers, and were thus excluded by \cite{butler_intracluster_2025}.} at $z=0.06$ as used in \cite{butler_intracluster_2025}. These are the most massive haloes in this snapshot, and span a mass range\footnote{Defined as the mass within a sphere of radius $R_\mathrm{178c}$ in which the average overdensity is 178 times the critical density, which corresponds more closely to the expectation from an analytic top-hat collapse model than the overdensity of 200 used in Hydrangea. Masses quoted in $M_\mathrm{178c}$ are therefore slightly larger than their equivalent $M_\mathrm{200c}$, but as our samples are too small for detailed mass matching this difference is irrelevant here.} of $M_{\rm 178c} = 1.18-3.71 \times 10^{14}\,{\rm M}_\odot$.

Halo finding in Horizon-AGN is done in a fundamentally different way from Hydrangea, using the \textsc{AdaptaHOP} \citep{Aubert2004MNRAS.352..376A, Tweed_2009} structure finder. This code identifies structures purely based on the local density associated with each particle, without consideration of gravitational boundedness. Excising satellites based on their \textsc{AdaptaHOP} membership therefore causes both artificial holes in the BCG+ICL and DM -- since particles that are physically part of the diffuse stellar halo but happen to be spatially coincident with a satellite galaxy are also cut out -- and prominent shell-like residuals around them, as particles that are physically part of satellites tend to be missed from their outskirts. Defining the Horizon-AGN BCG+ICL and DM halo based on the \textsc{AdaptaHOP} classification would therefore risk introducing significant systematic differences with respect to Hydrangea that are related not to actual differences in the simulations, but rather in their post-processing. To enable a direct comparison unaffected by differences from these different halo finders, we therefore apply the simplified satellite removal method described in Section \ref{sec:sat_removal_comparison} identically to both simulations. 

\begin{figure*}
	\includegraphics[width=2\columnwidth]{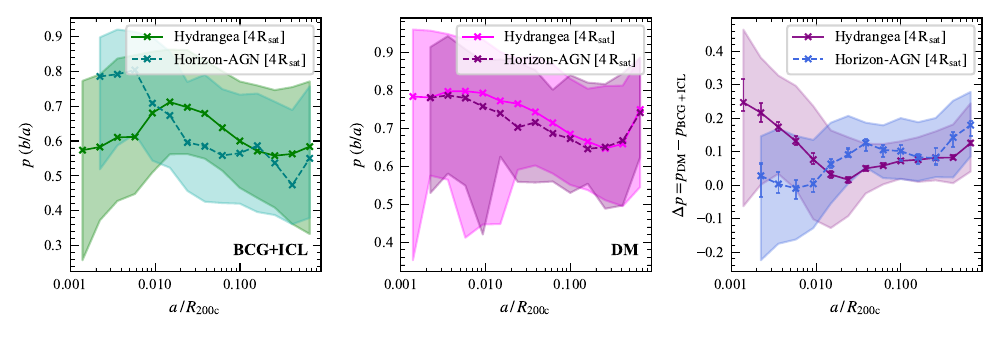}
    \caption{Comparison of projected BCG+ICL and DM axis ratios between the Hydrangea and Horizon-AGN simulations. In both, satellites are removed by excising particles within $4 r_\mathrm{sat}$ from identified subhaloes (see text). Left: the median axis ratios, $p=b/a$, of the BCG+ICL component in Hydrangea (green solid line) and Horizon-AGN (dashed blue line), as a function of $a/R_{\mathrm{200c}}$. Middle: the same for DM, with a solid magenta line for Hydrangea and a dashed purple one for Horizon-AGN. Right: median axis ratio difference for Hydrangea (purple solid line) and Horizon-AGN (blue dashed line). In all cases, lines represent medians, shaded regions the 1$\sigma$ scatter, and error bars (in the right panel) the 1$\sigma$ uncertainties on the median estimated via bootstrap resampling. While the two simulations agree broadly with each other for the BCG+ICL, and closely for the DM, the offset between both components is statistically different between the two models, at all radii.}
    \label{fig:Hydrangea and Horizon}
\end{figure*}

The comparison between Horizon-AGN and Hydrangea, using the simple $4 r_\mathrm{sat}$ satellite excision for both, is shown in Fig.~\ref{fig:Hydrangea and Horizon}; the layout is analogous to Fig.~\ref{fig:Mass split}. The left panel displays the median axis ratio profiles, $p=b/a$, of the BCG+ICL components in Hydrangea [$4\,r_\mathrm{sat}$] (solid green line) and Horizon-AGN [$4\,r_\mathrm{sat}$] (dashed blue line). In the BCG-dominated region we find that $p$ is significantly higher in Horizon-AGN [$4\,r_\mathrm{sat}$], reaching a maximum of $\approx 0.8$, so its BCGs are significantly rounder. This difference likely arises from the different treatment of feedback processes in the two simulations, in particular for AGN. In the ICL-dominated region the profiles are generally more similar, although the Horizon-AGN [$4\,r_\mathrm{sat}$] clusters have systematically more elliptical BCG+ICL, by up to $\approx$0.1. This suggests that the distribution of the ICL may have some dependence on the baryon physics implemented in the simulation, although it could also be an effect of more pronounced residuals in Horizon-AGN [$4\,r_\mathrm{sat}$] compared to Hydrangea [$4\,r_\mathrm{sat}$]\footnote{Recall that the excision aperture $r_\mathrm{low}$ for low-mass galaxies is based on the sizes of Horizon-AGN galaxies, whereas the EAGLE model of Hydrangea is explicitly calibrated against observational galaxy sizes so that these are, in general, smaller. In effect, satellites are therefore removed somewhat more fully in Hydrangea [$4\,r_\mathrm{sat}$].}. The DM profiles are overall similar in both median $p$ and scatter (middle panel of Fig.~\ref{fig:Hydrangea and Horizon}), particularly in the very inner regions ($a \gtrsim 0.004$\,$R_{\mathrm{200c}}$) and the outskirts ($a \gtrsim 0.3$\,$R_{\mathrm{200c}}$). At intermediate radii, the DM haloes in Hydrangea [$4\,r_\mathrm{sat}$] are slightly rounder, with a maximum discrepancy of $\approx 0.06$. As we would not expect significant differences in the modelling of gravity between the two simulations, this discrepancy might originate from an imperfect match between the two cluster samples. 

Our main concern here, the axis ratio differences ($\Delta p =p_{\mathrm{DM}} - p_{\mathrm{BCG+ICL}}$), are shown in the right panel of Fig.\,\ref{fig:Hydrangea and Horizon} for both simulations. In the BCG-dominated region, $\Delta p$ in Horizon-AGN [$4\,r_\mathrm{sat}$] is significantly lower than in Hydrangea [$4\,r_\mathrm{sat}$], with minimal overlap in the scatter. This is consistent with the offset seen in the BCG+ICL alone, and implies that we must treat our previous result of the relatively poor match between the BCG and DM shapes with caution. At ICL-dominated radii the profiles become more similar, but the median $\Delta p$ for Horizon-AGN [$4\,r_\mathrm{sat}$] is on average $\approx\! 0.03$ higher, reflecting a more elongated BCG+ICL component. Although there is substantial overlap between the scatter of the two simulations, the medians are discrepant at a $\approx\! 2\sigma$ level (estimated via bootstrapping) in most individual bins, except around $\sim 0.2$\,$R_{\mathrm{200c}}$. As discussed above, this offset might reflect differences in the (stellar) residuals rather than in the physical ICL between the two simulations. However, it is also possible that the discrepancy lies in the ICL itself, and therefore that the offset between ICL and DM is dependent, in detail, on the simulation model as well as the satellite excision method.

While these comparisons demonstrate that the axis ratio offset between the DM and BCG+ICL is sensitive in detail to both the satellite removal method used and the underlying simulation model, the effect is generally small compared to the magnitude of the offset itself, as shown in Fig.~\ref{fig:diff plot Hydrangea}. At ICL-dominated radii, the DM-ICL axis ratio offsets between Horizon-AGN [$4\,r_\mathrm{sat}$] and Hydrangea [$4\,r_\mathrm{sat}$] are $\approx\! 0.11$ and $\approx\! 0.08$, respectively, so both are qualitatively similar. Regardless of the exact choice of simulation, our results therefore confirm that the ICL is a good tracer of the DM halo shape.

\section{Comparison to the satellite galaxy distribution as a DM tracer}
\label{sec:other_tracers}
The distribution of satellite galaxies in clusters has frequently been used as an observational proxy for the shape of the underlying DM halo, although it is known to be biased \citep[e.g.,][]{Uitert2017,herle_unbiased_2025}. It is therefore of interest to test whether the BCG+ICL traces the halo shape better than the satellite population.

To quantify the shape of the galaxy distribution in each of our Hydrangea clusters, we select all subhaloes with stellar mass $M_\mathrm{star} \geq 10^{10}\,\mathrm{M}_\odot$ within $R_\mathrm{200c}$ of the  cluster, to mimic the magnitude limit of observational surveys \citep[e.g.][]{Rykoff2016,Benoist2025}. The shape tensor is then computed from all these galaxies, in 2D (projected along three orthogonal axes) for consistency with observations. The only difference from the particle-based method described in Section~\ref{defining_BCG+ICL} is that we only consider one single radial bin, from 0 to $r = R_\mathrm{200c}$, due to the orders of magnitude lower number of satellites than stellar or DM particles. We verified that attempting to measure shapes using all stellar particles associated with satellite galaxies across 30 radial bins leads to poor convergence at many radii, and to unstable, extremely elongated shapes in those bins where convergence is achieved. We therefore tested two single aperture approaches, treating the satellite galaxies as point tracers and using their constituent stellar particles. We find that although both approaches yield similar misalignment angles, the particle-based treatment results in a larger median DM-galaxy axis ratio offset and increased scatter. Combined with the fact that galaxies are typically treated as point tracers in observational estimates of the halo shape (e.g.~\citealt{wojtak2013, shin2018, zhou2023}), we therefore adopt the point-based approach. We also verified that adopting an all-particle DM definition, including both the smooth main halo and subhalo particles, in place of the main halo DM yields no significant difference in either $\Delta p$ or $\Delta \phi$. From the point-based approach, we obtain a single axis ratio $p_{\rm{gal}}$ and position angle $\phi_{\rm gal}$ for each cluster projection, and hence the offset of these values from $p_\mathrm{DM}$ and $\phi_\mathrm{DM}$, respectively. Because the latter vary with radius, the same is true for the differences even if $p_\mathrm{gal}$ and $\phi_\mathrm{gal}$ are constant.

\begin{figure}
	\includegraphics[width=\columnwidth]{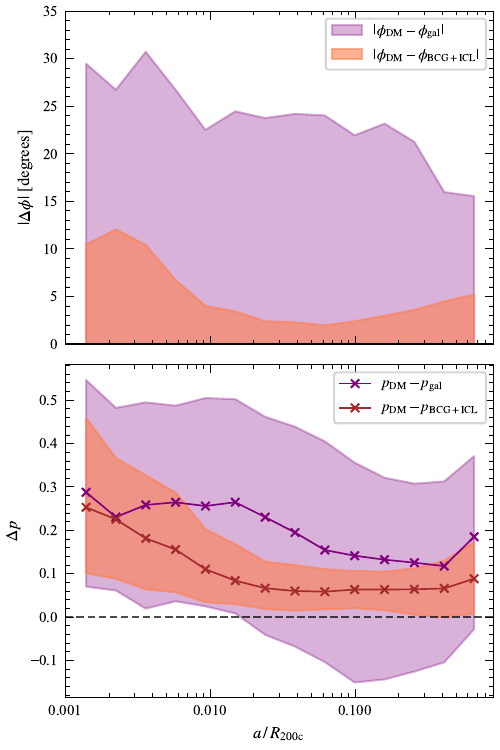}
    \caption{Comparison of ICL and satellite galaxies as tracers of the DM halo shape. Top: absolute differences in position angle between DM and the galaxy distribution ($|\phi_{\mathrm{DM}}-\phi_{\mathrm{gal}}|$; purple), and between the DM and BCG+ICL ($|\phi_{\mathrm{DM}}-\phi_{\mathrm{BCG+ICL}}|$; orange). Shaded regions show the 0-50$^\mathrm{th}$ percentile range, plotted against $a/R_{\mathrm{200c}}$. Bottom: the same, but for the difference in axis ratio $p$; solid lines show the median and shaded regions indicate the $1\sigma$ scatter and the dashed line indicates $\Delta p = 0$. In both orientation and axis ratio, the BCG+ICL traces the DM much more closely than satellites.}
    \label{fig:galaxies}
\end{figure}

The top panel of Fig.~\ref{fig:galaxies} compares the difference in position angles between DM and galaxies (purple) to that with respect to the BCG+ICL (orange), as shown in Fig.~\ref{fig:diff plot Hydrangea}. For clarity, only the median difference is shown for both. It is immediately obvious that the DM-galaxy alignment is much weaker than for the BCG+ICL, with typical differences of $\approx$15--30 degrees (highest in the centre), a factor $\approx$3 larger than for the BCG+ICL. However, the misalignment is still well below the 45 degree difference expected for a fully random distribution (especially in the cluster outskirts). In the absence of ICL measurements, the orientation of the satellite distribution can therefore still serve as a reasonable proxy for that of the outer DM halo with a precision that is comparable to assuming the orientation of the BCG, given the typical $\sim$10 degree misalignment between the BCG and ICL measured in {\it Euclid} Q1 (Kolcu et al., in prep.), and a $\sim$5 degree offset between the ICL and DM in the outskirts.

In the bottom panel of Fig.~\ref{fig:galaxies} we plot the axis ratio differences, again comparing the offset between DM and galaxies (purple) to that with respect to the BCG+ICL (orange). The two are less strongly discrepant than for the major axis position angle, but 
$\Delta p$ of the satellite galaxy distribution is still significantly higher than for the BCG+ICL, across all radii except in the very centre (by $\lesssim$0.15 in the median). This implies that the galaxy distribution, when used as a DM tracer, is biased even more towards more elliptical shapes than the BCG+ICL. The scatter is also much larger, up to $\approx 0.5$ throughout the BCG region and still up to $\approx$0.3 near the cluster edge; notably a significant minority of clusters also have a negative $\Delta p_\mathrm{gal}$, in other words their galaxy distribution is rounder than the DM halo. Since galaxies are much more similar to DM in terms of their kinematics than the BCG+ICL \citep{butler_intracluster_2025}, the poorer agreement for satellites than the BCG+ICL is likely due to increased sampling noise from the sparsity of galaxies (of order 100, compared to $\sim 10^4$ star particles per radial bin). We therefore conclude that the BCG+ICL provides a more reliable tracer of the shape of the DM halo than the galaxy distribution, at least when using the shape tensor to quantify the latter.

\section{Discussion}
\label{sec:discussion}
The main overarching conclusion of our analysis presented above is that the BCG+ICL provides a robust tracer of the DM shape, with only a small offset in axis ratio and position angle that remains consistent across 3D and 2D measurements and cluster properties. We now investigate the physical origins of this behaviour, and consider how this close correspondence can be confirmed and leveraged in future observational studies.

A plausible explanation for the close axis alignment is that both the ICL and DM in clusters result from the accretion of matter along the same preferred directions, as traced by large-scale cosmic filaments \citep{kuchner_mapping_2020, 2022ApJS, vurm2023A&A}. The subsequent evolution of both components is governed by collisionless dynamics in the same large-scale gravitational potential, so that this initial alignment is preserved. A close alignment in the orientation has also been reported by \citet{gonzalez_halo_2021}, who like us measured the shapes of ICL and DM in simulations via a shape tensor method. However, they only computed three integrated quantities for each cluster (all particles within $R_\mathrm{1000c}$, $R_\mathrm{500c}$, and $R_\mathrm{200c}$) rather than a full profile using ellipsoidal shells as we have done here. They find a typical misalignment of $\lesssim$10–20\,degrees, which is in excellent agreement with our results considering that integrated measurements might, in effect, give slightly different radial weighting to the two components.

With this close agreement in position angles, the ICL can serve as a direct observational proxy for the orientation of the underlying DM halo. This has important implications for weak lensing analyses that typically stack shear maps from many clusters to enhance the intrinsically weak lensing signal (e.g.~\citealt{Gonzalez2021}). Stacking in random orientations will inevitably average out any intrinsic ellipticity and result in an artificially circular shape that, in turn, may lead to a bias in the inferred mass and/or concentration of the halo (e.g.~\citealt{bahe_mock_2012}). It would therefore be preferable to align the individual clusters along their major axes before stacking. A commonly used proxy for this (unknown) DM halo orientation is the distribution of satellite galaxies. However, as we have shown in Sect.~\ref{sec:other_tracers}, the major axis of the satellite galaxies is only a relatively weak proxy for that of the DM, with typical deviations of $\sim 20$\,degrees. As the ICL is much more closely aligned with the DM halo, using its major axis to align the shear maps prior to stacking is a promising route to reduce systematic uncertainties and obtain more accurate ellipticity measurements of the DM haloes.

In addition to their near-perfect alignment, we also find close agreement between the ICL and DM axis ratios. This is consistent with previous studies that have established a connection between the isophotal contours of the ICL and the total mass distribution in clusters \citep{montes_intracluster_2019, alonsoasensio_intracluster_2020, yoo_comparison_2022}. Our findings complement these isocontour-based works by yielding 3D shape measurements, and by quantifying the difference in shape and orientation in terms of relative roundness. The fact that we find both the position angle and axis ratio difference between the ICL and DM to be small suggests that the two components are physically associated: it is consistent with both having been accreted at broadly similar epochs during cluster assembly so that both have had comparable time to (partially) relax.

In spite of this agreement, the ICL and DM do not trace each other perfectly. The axis ratios have a typical offset of $\Delta p \approx 0.07$, with the DM being slightly more spherical. However, its magnitude depends, in detail, on both the satellite removal technique and the galaxy formation model implemented in the simulations. The number quoted above can therefore not be applied directly to observations (or other simulations that were processed in a different way) to infer a more precise DM axis ratio from the ICL, rather than simply assuming they are equal. A slightly rounder shape of DM haloes compared to the ICL has also been reported by \citet{gonzalez_halo_2021} who report a difference of up to $\approx$15 per cent between the two (albeit using isocontours), in good quantitative agreement with our results.

There are several plausible explanations for the ICL being marginally more elongated than the DM halo of the same cluster. One is that the DM in infalling subhaloes may be stripped and incorporated into the smooth main halo on shorter timescales than the stars, allowing it more time to relax into a rounder shape. This interpretation is supported by recent work of \citet{martin2026}, who show that DM is, on average, stripped earlier than stars and occupies higher orbital energies. Alternatively, the axis ratio difference between the ICL and DM could reflect a slightly different origin of the two components: while the ICL originates primarily from relatively massive satellite galaxies that preferentially fall in along the cluster's dominant filaments, the DM includes a contribution from very low-mass haloes (effectively smooth accretion) whose infall may be more isotropic. Some memory of these different origins may persist to the present day in the form of a slightly more anisotropic (i.e.~elongated) distribution of ICL compared to the DM. Finally, tidal stripping of stars into the ICL is thought to be most effective for those galaxies that are on highly radial orbits and hence pass close to the BCG where they experience the strongest tidal forces. This might bias the ICL towards more elongated shapes, whereas the majority of DM, less tightly bound to a galaxy, can be stripped more evenly across the satellite population and hence produce a rounder shape \citep{butler_intracluster_2025}. Direct tests of these processes could in principle be carried out by tracing particles across snapshots to compare the accretion histories of the ICL and DM, and to examine how the axis ratios and position angles of each component evolve with time. Related approaches have been explored by \citet{yoo2024ApJ...965..145Y}, who examine the evolution of the spatial similarity between the BCG+ICL and DM across snapshots.

In the near future, high-precision measurements of the DM and ICL shapes of statistical cluster samples (through stacked weak lensing and photometry analyses, respectively) will make it possible to directly test the connection between the two in the real Universe. The upcoming \textit{Euclid} data releases will provide an ideal data set for such a comparison. Our simulation predictions provide a quantitative estimate of the expected offset (at least with our specific satellite removal method) and also suggest that it can be reliably measured at relatively small cluster-centric distance ($\sim\!0.1 R_\mathrm{200}$), without the need to probe the faint outer parts of the cluster where the signal-to-noise would drop precipitously.

The non-zero offset in axis ratio between DM and ICL notwithstanding, to first order we find it to be close to zero so that our results demonstrate a clear potential of using the ICL as an observable DM tracer. Observational measurements of the ICL shape can therefore provide a powerful method to discriminate between competing DM models, provided their predictions differ by more than the intrinsic uncertainty imposed by the (slight) mismatch between DM and ICL as discussed above. The recent work by \citet{gonzalez_cluster_2024} stacked the weak lensing signals from large cluster samples to recover their average halo ellipticity, and used this method to compare halo shapes in simulations with CDM and SIDM. Our results suggest that similar constraints can be obtained for \emph{individual} clusters by measuring ICL shapes in \textit{Euclid} data.

A second opportunity that follows from the close correspondence between the ICL and DM axis ratios is to use the former to rank clusters by DM halo ellipticity. As we have shown, the offset between the ICL and DM axis ratios has smaller scatter ($\sigma_{p} \approx 0.06$ than the typical spread in the DM axis ratios themselves ($\sigma_{p_{\rm DM}} \approx 0.27$). On average, the clusters with the most elliptical ICL components must therefore also host the most elliptical DM haloes, and vice versa. Splitting a large cluster sample into subsets of different ICL axis ratio would therefore provide a feasible observational approach to investigating the dependence of DM halo shape on cluster properties such as mass, dynamical state, or concentration.

\section{Conclusions}
\label{sec:conclusions}

Understanding the shapes of dark matter haloes in galaxy clusters is important for probing their assembly history and the nature of DM. However, measurements of these shape are challenging and have traditionally relied on tracers with significant inherent limitations, such as X-ray emission or satellite galaxy distributions. We therefore investigated whether the ICL can serve as a reliable tracer of the shape of the underlying DM halo in galaxy clusters. Using clusters from the Hydrangea simulations, we measured the 3D and projected shapes of both components using the shape tensor computed in concentric ellipsoidal shells, treating the BCG and ICL as one continuous component (`BCG+ICL'). We further compared these shapes to those obtained using isophotal fitting to assess how well observational methods capture the true morphology.

Our main results are summarised as follows:
\begin{enumerate}
    \item The BCG+ICL and DM are closely aligned with each other. The angle between their respective major or minor axes is typically $\lesssim$10\,degrees, and $\lesssim$3 degrees at intermediate radii ($a \sim 0.1\,R_\mathrm{200c}$). This indicates that both components share a common orientation imposed by the global gravitational potential and the preferred directions of anisotropic infall (Fig.~\ref{fig:3D results}, left column).
    
    \item The BCG+ICL and the DM have almost the same axis ratios, with the DM being slightly rounder (Fig.~\ref{fig:3D results}, right column). The offset is greatest near the centre (0.2--0.3), and then decreases with radius, reaching a plateau of $\approx$0.07 at ICL-dominated radii ($a > 0.05\,R_\mathrm{200c}$).
    
    \item Both the orientation and axis ratios are consistent between 3D and 2D measurements (Fig.~\ref{fig:diff plot Hydrangea}), implying negligible contamination from projection effects. The BCG+ICL shape is also insensitive to the measurement method; we obtain near-identical results from fitting ellipses to isophotes in mock images (Fig.~\ref{fig:isophotal_results}). 
    
    \item While both the BCG+ICL and DM axis ratios depend on cluster mass, the offset between the two components does not (Fig.\,\ref{fig:Mass split}). Our results are therefore applicable to the full cluster population, including the rare massive ones that are most commonly targeted by observations.
    
    \item The axis ratio offset between the BCG+ICL and DM is sensitive to both the method used to remove satellites (Fig.~\ref{fig:Hydrangea Re Method}), and the simulations employed (Fig.~\ref{fig:Hydrangea and Horizon}). The latter may imply a small sensitivity to uncertain galaxy formation physics.

    \item The BCG+ICL traces the DM halo shape more faithfully than the satellite galaxies, in both axis ratio and position angle. Satellites have a stronger bias towards more elliptical axis ratios, accompanied by larger scatter of up to 0.5. This is likely due to the sparse sampling of the halo by the satellites (Fig.~\ref{fig:galaxies}).
    
\end{enumerate}

The overarching conclusion emerging from these separate results is that the ICL closely traces the underlying DM halo in both axis ratio and orientation, outperforming the satellite galaxy distribution as a DM shape proxy. As shown by pioneering observations prior to {\it Euclid} \citep{montes_intracluster_2019, kluge_structure_2020, cha_high-caliber_2025}, measuring these properties is now a realistic possibility that will soon be applicable to many thousands of galaxy clusters. At the same time, simultaneous ICL and weak lensing measurements of the same clusters will offer the possibility to confirm the small, systematic offset between the axis ratios of the ICL and DM. 

\section*{Acknowledgements}
AF, NAH, TK, and JB thank the Leverhulme Trust for support through a Research Leadership Award. YMB acknowledges support from UK Research and Innovation through a Future Leaders Fellowship (grant agreement MR/X035166/1) and financial support from the Swiss National Science Foundation (SNSF) under project ``Galaxy evolution in the cosmic web''  (200021\_213076). GM and NAH acknowledge support from the UK STFC under grant ST/X000982/1. MM acknowledges support from grant RYC2022-036949-I financed by the MICIU/AEI/10.13039/501100011033 and by ESF+, grant PID2024-158845NB-I00, financed by MICIU/AEI/10.13039/501100011033 and by ERDF, EU, and program Unidad de Excelencia Mar\'{i}a de Maeztu CEX2020-001058-M. This research was supported by the International Space Science Institute (ISSI) in Bern, through ISSI International Team project \#23-577. This work used the DiRAC@Durham facility managed by the Institute for Computational Cosmology on behalf of the STFC DiRAC HPC Facility (www.dirac.ac.uk). The equipment was funded by BEIS capital funding via STFC capital grants ST/K00042X/1, ST/P002293/1, ST/R002371/1 and ST/S002502/1, Durham University and STFC operations grant ST/R000832/1. DiRAC is part of the National e-Infrastructure. The Hydrangea simulations were in part performed on the German federal maximum performance computer ‘HazelHen’ at the maximum performance computing centre Stuttgart (HLRS), under project GCS-HYDA / ID 44067, financed through the large-scale project ‘Hydrangea’ of the Gauss Center for Supercomputing. Further simulations were performed at the Max Planck Computing and Data Facility (MPCDF) in Garching, Germany.


\section*{Data Availability}
The Hydrangea simulations are publicly available at \url{https://ftp.strw.leidenuniv.nl/bahe/Hydrangea}. The raw data products of the Horizon-AGN simulation are available upon reasonable request at \url{https://www.horizon-simulation.org/}. High-level data generated for this work will be made available upon reasonable request to the corresponding author.



\bibliographystyle{mnras}
\bibliography{references_updated, alternative} 



\appendix

\section{Quantification of shape measurement uncertainties}
\label{app: error grids}
\begin{figure*}
    \centering
    \includegraphics[width=1\linewidth]{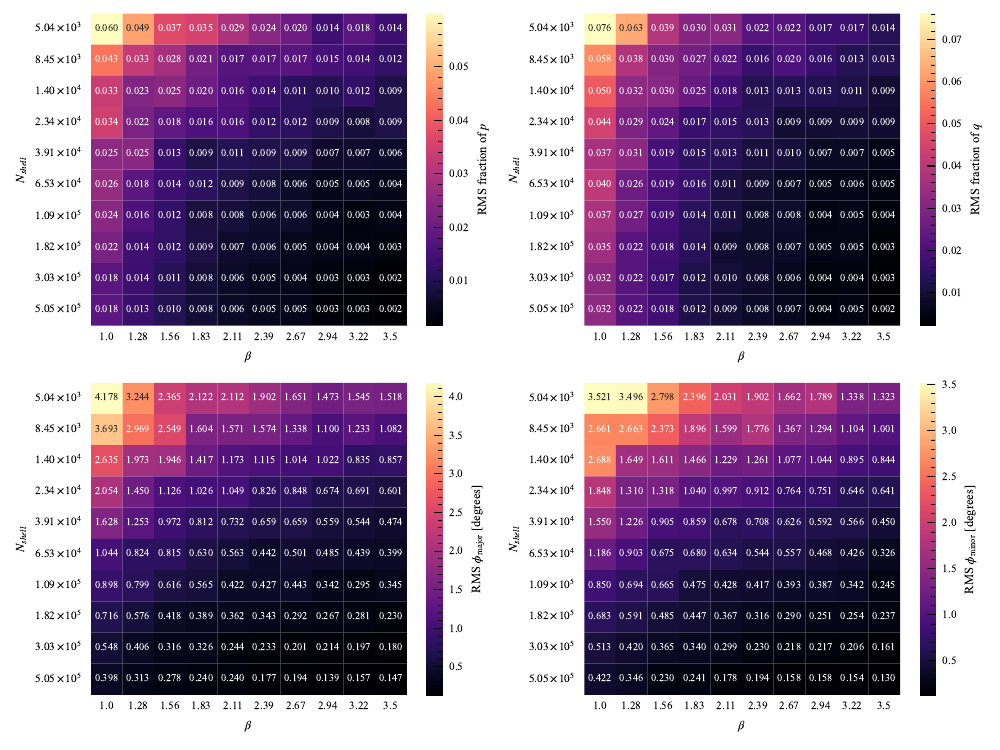}
    \caption{3D shape measurement uncertainties from mock ellipsoids. The four panels show the fractional (top) or absolute (bottom) uncertainties on the axis ratios $p=b/a$ (top left) and $q=c/a$ (top right), respectively, as well as the major axis (bottom left) and minor axis position angles (bottom right). They are provided for different numbers of particles per shell ($N_{\rm shell}$; along the y axis) and the slope of the density profile ($\beta$; along the x axis). For each combination, the quoted value is the fractional root-mean-square deviation from the true input value, across 50 runs that vary only in their random seed. The uncertainties are generally negligible ($\lesssim$ 1 per cent or 1 degree), except for the shallowest profiles sampled with fewer than $\sim 3 \times 10^4$ particles.}
    \label{fig:3D_error_grid}
\end{figure*}

We present here the full results of our tests to quantify the uncertainty in our shape measurements through the analysis of analytic mock ellipsoids (see Sect.~\ref{sec: sims and BCG+ICL definition}). We generate 3D ellipsoids of semi-major axis length of 1000\,kpc, axis ratios $p=0.7$ and $q=0.5$, and orientations defined by by Euler angles $(\alpha, \beta, \gamma) = (30^\circ, 30^\circ, 30^\circ)$. We perform shape measurements on the particles within a shell at $r=250$\,kpc with a thickness of $\Delta r=50$\,kpc. We vary both the number of particles in each shell, $N_{\rm shell}$, and the density profile slope, $\beta$, over the ranges $10^{3.7}<N_{\rm shell}<10^{5.7}$ and $1.0<\beta<3.5$. For each point in this parameter space, 50 realisations are produced by changing only the random seed, which changes the locations of individual particles. The shape tensor (Eqn.\,\ref{eqn:tensor element}) is then evaluated to recover the axis ratios and position angles. The fractional root-mean-square of the deviation of the measured values from the input position angles and axis ratios is displayed in Fig.~\ref{fig:3D_error_grid} as our corresponding uncertainty estimates. The top left and right panels correspond to the uncertainties on $p$ and $q$, respectively, and the bottom left and right to the uncertainties on position angles of the major and minor axes, respectively. Both $N_{\rm shell}$ and $\beta$ affect the uncertainty to a similar extent: higher particle numbers or steeper density slopes result in lower uncertainties. Our radial bin selection applied to the Hydrangea simulations results in typical ranges (16$^\mathrm{th}$-84$\mathrm{th}$ percentile) of $N_{\rm shell} \approx 10^{4.2}-10^{4.8}$ and $\beta \approx 2.6-3.4$ for the ICL, and $N_{\rm shell} \approx 10^{3.5}-10^{6.0}$ and $\beta \approx 1.1-2.2$ for the DM. These result in uncertainties of $\approx 0.5-1.2$ per cent in ICL axis ratios and $\sim0.4-1.0$\,degrees in position angles, and $\approx 0.6-5.7$ per cent in DM axis ratios and $\approx 0.2-3.9$\,degrees in position angles.

We additionally test the impact of an offset between the assumed reference centre of the shape measurement method and the true mass distribution. The cluster centres in Hydrangea are defined as the location of the particle with the lowest potential, which always corresponds with the centre of the BCG. Consequently, only the DM component can be offset from the cluster centre. Using mock 3D ellipsoids generated as above with the median Hydrangea DM  $N_{\rm shell}$ and $\beta$ values, we introduced spatial offsets ranging from 1 to 10\,kpc and measure shapes in 30 logarithmically spaced radial bins between 3 and 1000\,kpc. Offsets mainly affect the innermost radii, while for realistic offsets (up to 5\,kpc; \citealt{Harvey2019}), beyond a radius of $\sim40$\,kpc the absolute differences between the input and measured values remain below 0.022. Although this value is statistically significant compared to the uncertainties, it is substantially smaller than the axis ratios of the BCG+ICL and DM, as well as the axis ratio differences between them. This indicates that, in the ICL-dominated region, the effect of realistic offsets on our conclusions is negligible.

\begin{figure*}
    \centering
    \includegraphics[width=1\linewidth]{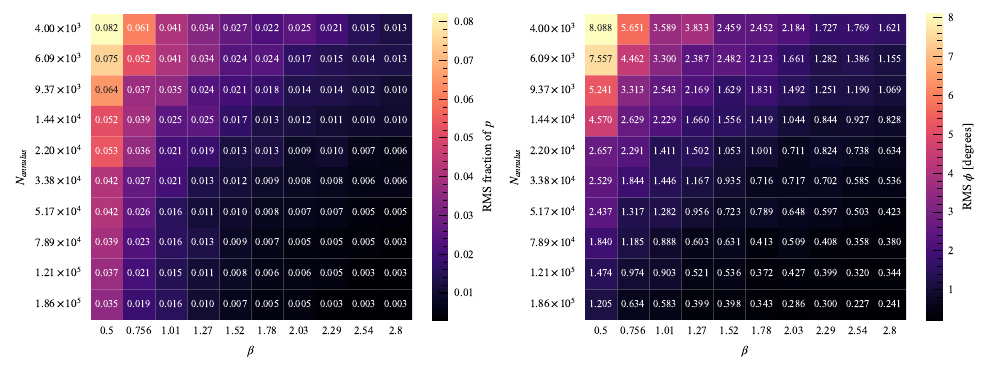}
    \caption{Same as Fig.~\ref{fig:3D_error_grid}, but for 2D. The left-hand panel shows the uncertainty on the axis ratio, the right-hand panel that on the position angle. As for 3D, they are generally negligible.}
    \label{fig:2D_error_grid}
\end{figure*}

We also estimate the uncertainties in our 2D shape measurements. To achieve this, we generate mock ellipses of axis ratio $p=0.75$ and position angle $\phi=30$\,degrees, adopting the same semi-major axis length of 1000\,kpc. We vary $N_{\rm shell}$ and $\beta$ across the ranges $10^{3.6}<N_{\rm shell}<10^{5.3}$ and $0.5<\beta<2.8$, respectively. In a process analogous to the 3D case, we produce 50 ellipses for each combination and measure the shapes of the particles within an annulus at $r=250$\,kpc with width $\Delta r=50$\,kpc. The fractional root-mean-square of the deviation from the true input values is shown in Fig.~\ref{fig:2D_error_grid} which used to estimate the uncertainties on 2D shape measurements. The left and right panels show the uncertainties on $p$ and $\phi$, respectively. We observe a similar degeneracy to that in the 3D case.

These uncertainties on individual cluster measurements are smaller than the cluster-to-cluster scatter and are therefore not shown in the figures of the main paper.


\bsp	
\label{lastpage}
\end{document}